\documentclass[11pt,a4wide]{article}

\usepackage{hyperref}
\usepackage{amsmath}
\usepackage{amsfonts}
\usepackage{amsbsy}
\usepackage{epsfig}
\usepackage{soul}
\usepackage{latexsym}
\usepackage{xcolor}
\input amssym.def
\input amssym.tex

\advance \topmargin by -\headheight
\advance \topmargin by -\headsep
\evensidemargin \oddsidemargin
\advance\hoffset by -3mm  
\advance\voffset by  8mm  
\def\bbbc{{\mathchoice {\setbox0=\hbox{$\displaystyle\rm C$}\hbox{\hbox
to0pt{\kern0.4\wd0\vrule height0.9\ht0\hss}\box0}}
{\setbox0=\hbox{$\textstyle\rm C$}\hbox{\hbox
to0pt{\kern0.4\wd0\vrule height0.9\ht0\hss}\box0}}
{\setbox0=\hbox{$\scriptstyle\rm C$}\hbox{\hbox
to0pt{\kern0.4\wd0\vrule height0.9\ht0\hss}\box0}}
{\setbox0=\hbox{$\scriptscriptstyle\rm C$}\hbox{\hbox
to0pt{\kern0.4\wd0\vrule height0.9\ht0\hss}\box0}}}}

\makeatletter
\@addtoreset{equation}{section}
\makeatother

\begin{document}
\title{{\Large \bf{Higher-Dimensional Vacuum Einstein Equations: Symmetry, New Solutions, and Ricci Solitons}}}
\author{
M M Akbar$^{1,}$\footnote{E-mail: akbar@utdallas.edu}
\ \,\,\&\,
M Self$^{2,}$\footnote{E-mail: mmself@ucdavis.edu}
\\
\\ {$^{1}${Department of Mathematical Sciences}}
\\ {University of Texas at Dallas}
\\ {Richardson, TX 75080, USA}
\\
\\ $^{2}${Department of Physics}
\\ {University of California at Davis}
\\ {Davis, CA, 95616}}
\date{\today}
\maketitle
\begin{abstract}
We show that the system of vacuum Einstein equations (i.e., Ricci-flat metrics) with two hypersurface-orthogonal, commuting Killing vector fields in $d \ge 5$ dimensions is invariant under the action of a one-parameter Lie group, and the group action on any metric can be expressed in a closed, universal form. This enables the generation of a one-parameter family of solutions from any given ``seed" solution of the system without solving additional equations, as well as one-parameter families of local steady Ricci solitons. This extends the Lie point symmetry in four dimensions, found earlier in \cite{AkbarMacCallum} for axisymmetric static vacuum systems, and provides the first example of solution generation in higher-dimensional vacuum Einstein equations that can be realized purely algebraically.
\end{abstract}
\section{Introduction}
Despite Einstein's initial concern that his equations might be unsolvable, numerous exact solutions were discovered during his lifetime, thanks to a talented generation of relativists who leveraged the mathematical techniques available at the time. Much of our understanding of the nonlinear nature of Einstein’s equations derives from these solutions. They underpin many key results in astrophysics, cosmology, and quantum gravity, and continue to guide our exploration of more general scenarios. Today, we have an impressive collection of exact solutions in four dimensions, which have been thoroughly classified and cataloged \cite{ExactSol2003, Griffiths, MacCallum2006}.

About two decades before the arrival of general relativity, Sophus Lie developed what would come to be known as Lie groups to study differential equations systematically. However, while Lie groups gained prominence in mathematics and physics, their original application to differential equations remained largely overlooked for about half a century (see, for example, \cite{Olver, Stephani}). This period overlapped with the most active era of constructing new solutions in general relativity, which could have greatly benefited from the application of Lie's methods to Einstein's equations. This was ironic, given that relativists had been formally using Lie groups for external spacetime symmetries from the very beginning. Lie's original ideas were revived and expanded in the early 1960s, starting with the work of Garrett Birkhoff\footnote{Garrett Birkhoff was the son of George Birkhoff, the originator of the eponymous theorem in general relativity. He passed away in 1944, prior to his son's interest in Lie groups.}. However, these ideas did not gain traction within the relativity community for about another decade.

There were ``generation techniques" in general relativity that were an exception. In the 1950s, Buchdahl demonstrated how one could obtain a static Ricci-flat metric “without solving the field equations” from another, and Ehlers showed how to derive a vacuum stationary solution from a vacuum static metric by solving an associated set of equations—both of which appealed to the hidden symmetries of the associated system of Einstein equations \cite{Buchdahl1954, Buchdahl1956, Buchdahl1959, Ehlers, Buchdahl1978}. Later, in the early 1970s, Geroch provided an algorithm to create an infinite-parameter family from any stationary axisymmetric metric \cite{Geroch1, Geroch2}. The idea behind such generation techniques was to find solutions that were difficult to obtain by directly integrating the field equations. However, it was not until the late 1970s that a vigorous study of stationary axisymmetric systems was conducted (see \cite{Hoenselaers:1985qk} and \cite[Chapter 34]{ExactSol2003}), fueled by the now well-developed symmetry methods for PDE systems in other areas of mathematics. A deep understanding of the symmetries of vacuum and matter-coupled rotating axisymmetric systems and their corresponding solution spaces was obtained (see, for example, \cite[Chapter 34]{ExactSol2003}). However, this understanding did not lead to new generation techniques where new solutions could be generated by simpler means without solving the original Einstein's equations or a set of equations involving lengthy steps.

In the special case of axisymmetric static system (i.e., when the two Killing fields are hypersurface-orthogonal and commuting), one can derive a distinct one-parameter family of metrics from the scaling invariance of Einstein's equations in Weyl coordinates \cite{ChngMannStelea}. Another well-known method for the same system was due to Ernst, which involves solving a relatively simple system of two differential equations to obtain a family of one-parameter static axisymmetric metrics from a given one \cite{Ernst1978, Ernst1979}. Both of these methods owe their existence to the Lie point symmetries of the field equations and are unique cases within the larger picture of Lie point symmetries of the system. More recently, another explicit symmetry was identified for this system \cite{AkbarMacCallum}. This symmetry, just like the scaling symmetry, can be used algebraically to generate a one-parameter family of static solutions from a seed solution without solving any intermediate equations.

This paper shows that in the presence of two hypersurface-orthogonal Killing vector fields the above Lie symmetry continues to exist in higher dimensions and that one can obtain a one-parameter family of vacuum solutions from any solution of this system without solving any equations. Over the past few decades, there has been significant interest in higher-dimensional gravity, both in vacuum and with matter fields, as these arise in string theory and other fundamental theories. However, there has not been a focused effort to understand the symmetries of higher-dimensional Einstein equations. We hope this work will deepen interest in the symmetries of Einstein's equations in higher dimensions.

\section{Commuting Orthogonal KVFs and Einstein Equations in Vacuum}
We will consider solutions of the vacuum Einstein equations in \(d \geq 4\) dimensions that possess two commuting, hypersurface-orthogonal Killing vector fields  --- \(\partial/\partial x\) and \(\partial/\partial y\) --- i.e., metrics that can be locally expressed in the following doubly-warped product form
\begin{equation}\label{Base Metric}
ds^2 = \pm g_{11} \, dx^2 \pm g_{22} \, dy^2 + g_{ij} \, dz^i \, dz^j,
\end{equation}
where \(i, j \in \{3, \ldots, d\}\), and all the metric components are functions of \(z^i\).

\subsection{Four Dimensions and Three Solution-Generation Techniques}
In four dimensions, vacuum metrics of the form (\ref{Base Metric}) arise in static axisymmetric systems and cylindrical gravitational wave systems. If, instead of (\ref{Base Metric}), one uses Weyl coordinates,
\begin{equation}
ds^2 = -e^{2u(\rho, z)} \, dt^2 + e^{-2u(\rho, z)} \left[ e^{2k(\rho, z)} (d\rho^2 + dz^2) + \rho^2 \, d\phi^2 \right], \label{basic1}
\end{equation}
the set of vacuum Einstein equations decomposes into a linear part,
\begin{equation}
u_{,\rho\rho} + \frac{u_{,\rho}}{\rho} + u_{,zz} = 0, \label{laplace}
\end{equation}
which is the axially symmetric Laplace equation in three-dimensional Euclidean space, and a nonlinear part,
\begin{equation}
k_{,\rho} = \rho \left[ \left( u_{,\rho} \right)^2 - \left( u_{,z} \right)^2 \right], \quad k_{,z} = 2 \rho \, u_{,\rho} \, u_{,z}, \label{combined1}
\end{equation}
which is fully determined by the solution of (\ref{laplace}). Such a splitting is not possible if one uses the more general form (\ref{Base Metric}) or any other choice of coordinates. With this setup, any solution of the system \((u(\rho, z), k(\rho, z))\) is uniquely determined by the harmonic function \(u(\rho, z)\) alone, often referred to as the axially symmetric ``Newtonian potential." However, note that one would still need \(k(\rho, z)\) to write down the metric. Being able to work out \(k(\rho, z)\) easily is a key aspect of any solution-generation technique.

One can readily see that the scaling
\begin{equation}
(u, k) \rightarrow (\beta u, \beta^2 k) \label{symm11}
\end{equation}
leaves the system (\ref{laplace})-(\ref{combined1}) invariant. Thus, this can be used to generate a one-parameter family without additional calculations from any seed solution \((u, k)\). Despite its simplicity, the resulting one-parameter metrics represent distinct geometries for \(\beta \neq 1\). Note that this simple scaling symmetry in Weyl coordinates does not translate straightforwardly to other coordinates. The Einstein equations for the general warped product form (\ref{Base Metric}) or any other coordinates will do a good job in hiding this.

The other explicit Lie point symmetry of the system (\ref{laplace})-(\ref{combined1}), as mentioned in the introduction, is the following transformation \cite{AkbarMacCallum}:
\begin{equation}\label{symakmac}
(u, k) \rightarrow (u + \alpha \ln \rho, k + 2\alpha u + \alpha^2 \ln \rho), \quad \alpha \in (-\infty, \infty).
\end{equation}
This symmetry also allows us to generate new one-parameter solutions effortlessly. However, unlike the scaling symmetry, this can, fortunately, be expressed succinctly in the warped product form:
\begin{eqnarray}
ds^2 = \pm (g_{22})^{\gamma} (g_{11})^{\gamma} g_{11} \, dx^2 &\pm& (g_{22})^{-\gamma} (g_{11})^{-\gamma} g_{22} \, dy^2 \nonumber \\
&+& (g_{22})^{\gamma(\gamma - 1)} (g_{11})^{\gamma(\gamma + 1)} g_{ij} \, dz^i \, dz^j, \quad \gamma \in (-\infty, \infty). \label{warpedformorggen}
\end{eqnarray}
These two symmetries, along with Ernst's prescription mentioned in the introduction, are the three known generation techniques available for the system (\ref{laplace})-(\ref{combined1}). For all of these, the group properties can easily be verified.

The Lie point symmetry (\ref{symakmac}) traces its origin back to the pioneering discrete transformation discovered by Buchdahl in the 1950s, as mentioned in the introduction. Buchdahl demonstrated how one could obtain a single Ricci-flat metric from another without solving any equations, in the presence of a hypersurface-orthogonal Killing vector field, which he referred to as a ``static" coordinate. He called the resulting metric the ``reciprocal" of the original. Later, he applied this transformation successively to both ``static" coordinates of the four-dimensional flat metric written in polar coordinates and observed that the discrete parameter could be extended to a continuous one. However, the resulting Ricci-flat metrics were already well-known solutions. Perhaps due to this outcome, or for reasons unknown to us, Buchdahl did not explore two hypersurface-orthogonal Killing vector fields separately in his subsequent work, even though the symmetries of the Einstein equations for axisymmetric vacuum systems were being vigorously studied at this time.\footnote{On the other hand, this study --- despite its sophistication --- did not identify the explicit symmetry of the static system (\ref{symakmac}).}

\subsection{Continuous Symmetry in Higher Dimensions}
With only two Killing vector fields, no coordinate system is known in higher dimensions in which Einstein's equations separate into a linear part and a nonlinear part. This is possible if there are \((d-2)\) hypersurface-orthogonal Killing vector fields, which will always be more than two in dimensions higher than four \cite{EmparanReallWeyl}. Thus, there will be multiple choices of two commuting hypersurface-orthogonal Killing vector fields, and the metric can be written in the form of (2.1). In addition to adding more flat directions, one can also consider generalizing the vacuum axisymmetric system by replacing the axial \(U(1)\) symmetry with \(SO(n)\) symmetry \cite{CharmousisGregory}. With, for example, the canonical metric on $SO(n)$ this will also allow for a commuting pair, and the metric can be cast in \eqref{Base Metric}.

In the following, we extend the symmetry \eqref{symakmac} of the four-dimensional vacuum Einstein equations to higher dimensions and seek analogs of \eqref{warpedformorggen}. We use Buchdahl's discrete transformation to construct the continuous symmetry from two sets of such transformations based on the two Killing vector fields. This approach obviates the need for a direct analysis of the Einstein equations for symmetry, which would be quite complicated in the absence of a coordinate system in higher dimensions, like the Weyl coordinates in four dimensions, when the number of Killing vector fields is strictly two.

Following Buchdahl, for every Ricci-flat metric of the form \eqref{Base Metric}, we can obtain two distinct Ricci-flat metrics
\begin{equation}\label{Buchdahl x1}
ds^{2} = \pm g_{11}^{-1} \, dx^{2} \pm g_{11}^{\frac{2}{d-3}} g_{22} \, dy^{2} + g_{11}^{\frac{2}{d-3}} g_{ij} \, dz^{i} dz^{j},
\end{equation}
and
\begin{equation}\label{Buchdahl x2}
ds^{2} = \pm g_{11} g_{22}^{\frac{2}{d-3}} \, dx^{2} \pm g_{22}^{-1} \, dy^{2} + g_{22}^{\frac{2}{d-3}} g_{ij} \, dz^{i} dz^{j}.
\end{equation}
Both are ``reciprocal metrics" of \eqref{Base Metric}, with $x$ and $y$ used as static coordinates, respectively, and are distinct from one another.
We would like to examine the result of applying these transformations repeatedly. It is clear that each transformation is its own inverse, so alternating them in succession yields a nontrivial result. Note that, apart from a possible difference in signature, equations \eqref{Buchdahl x1} and \eqref{Buchdahl x2} are simple relabels of one another under $x \leftrightarrow y$.

Without loss of generality, we apply the transformation \eqref{Buchdahl x1} first. It is easy to see that the resulting metric after applying the two transformations alternately $2n$ times is qualitatively different from the metric obtained after $2n+1$ times (i.e., \eqref{Buchdahl x1} applied $n+1$ times and \eqref{Buchdahl x2} applied $n$ times). In the former case, one obtains (where $q \equiv \frac{2}{d-3}$)
\begin{equation}\label{Discrete Transformation Even}
ds^{2} = \pm g_{11}^{\mathcal{A}_{2n+1}(q)} g_{22}^{\mathcal{A}_{2n}(q)} \, dx^{2} \pm g_{11}^{-\mathcal{A}_{2n}(q)} g_{22}^{-\mathcal{A}_{2n-1}(q)} \, dy^{2} + g_{11}^{\mathcal{B}_{2n}(q)} g_{22}^{\mathcal{B}_{2n-1}(q)} g_{ij} \, dz^{i} dz^{j},
\end{equation}
where the sequences $\mathcal{A}_{n}$ and $\mathcal{B}_{n}$ are defined recursively below. In the latter case, one obtains
\begin{equation}\label{Discrete Transformation Odd}
ds^{2} = \pm g_{11}^{-\mathcal{A}_{2n+1}(q)} g_{22}^{-\mathcal{A}_{2n}(q)} \, dx^{2} \pm g_{11}^{\mathcal{A}_{2n+2}(q)} g_{22}^{\mathcal{A}_{2n+1}(q)} \, dy^{2} + g_{11}^{\mathcal{B}_{2n+1}(q)} g_{22}^{\mathcal{B}_{2n}(q)} g_{ij} \, dz^{i} dz^{j}.
\end{equation}
We will see below that the symmetries $\mathcal{A}_{n} = -\mathcal{A}_{-n}$ and $\mathcal{B}_{n} = \mathcal{B}_{-n-1}$ hold. With these in mind, it is possible to recognize that \eqref{Discrete Transformation Odd} and \eqref{Discrete Transformation Even} can be combined into a single form:
\begin{equation}\label{Discrete Transformation}
ds^{2} = \pm g_{11}^{\mathcal{A}_{n+1}(q)} g_{22}^{\mathcal{A}_{n}(q)} \, dx^{2} \pm g_{11}^{-\mathcal{A}_{n}(q)} g_{22}^{-\mathcal{A}_{n-1}(q)} \, dy^{2} + g_{11}^{\mathcal{B}_{n}(q)} g_{22}^{\mathcal{B}_{n-1}(q)} g_{ij} \, dz^{i} dz^{j},
\end{equation}
which applies for all $n \in \mathbb{Z}$. The apparent distinction between \eqref{Discrete Transformation Odd} and \eqref{Discrete Transformation Even} arises only from a relative sign flip in the exponents of the first two terms, which can be accounted for by changing the sign of the subscript in $\mathcal{A}_{n}(q)$.

We now explicitly write the recursion relations for the above sequences:
\begin{equation}\label{A def'n}
\mathcal{A}_{n}(q) = q \mathcal{A}_{n-1}(q) - \mathcal{A}_{n-2}(q)
\end{equation}
and
\begin{equation}\label{B def'n}
\mathcal{B}_{n}(q) = \mathcal{B}_{n-1}(q) + q \mathcal{A}_{n}(q),
\end{equation}
with the initial conditions $\mathcal{A}_{0} = \mathcal{B}_{0} = 0$ and $\mathcal{A}_{1} = 1$. These recursion relations can be obtained by repeatedly applying the transformations \eqref{Buchdahl x1} and \eqref{Buchdahl x2} in sequence and observing their effect on the metric. It is easy to see that, in four dimensions (i.e., for $q=2$), $\mathcal{A}_n (2) = n$ and $\mathcal{B}_n (2) = n(n + 1)$ satisfy these conditions.

We can write a generating function for $\mathcal{A}_{n}$ by multiplying each term in the above recursion relation by $t^{n}$ for some real parameter $t$ and summing from $n=0$ to $\infty$. A bit of algebra allows us to isolate:
\begin{equation}\label{A Gen f'n}
\sum_{n=0}^{\infty} \mathcal{A}_{n}(q) t^{n} = \frac{t}{1 - qt + t^{2}},
\end{equation}
and by expanding the right-hand side of \eqref{A Gen f'n} into a power series, we obtain
\begin{equation}\label{A formula}
\mathcal{A}_{n}(q) = \frac{r^{n} - r^{-n}}{r - r^{-1}},
\end{equation}
where
\begin{equation}\label{r def}
r \equiv \frac{q + \sqrt{q^{2} - 4}}{2},
\end{equation}
the larger root of the equation $t^2 - qt + 1 = 0$. This holds for all $d > 4$ and is clearly an odd function of $n$. For $d = 4$, we have $q = 2$, and the denominator in \eqref{A Gen f'n} has a double root, which makes $r = 1$. This gives $\mathcal{A}_n (d=4) = n$, as mentioned above, which is also an odd function. Equation \eqref{A formula} clearly admits a generalization to a sequence indexed by a continuous parameter, which we call ${\gamma}$.

For \(d > 4\), we have \(q = \frac{2}{d - 3} < 2\). The characteristic polynomial \(1 - q t + t^2\) then has complex roots, which can be expressed as \(r = e^{i \theta}\) and \(\overline{r} = e^{-i \theta}\), with \(\theta = \cos^{-1} \left(\frac{q}{2}\right)\). The appearance of complex roots specifically for \(q < 2\) leads to an expression for \(\mathcal{A}_{\gamma}(d)\) in terms of trigonometric functions. Using these roots and expanding the generating function, we find
\begin{equation}\label{A trigonometric}
\mathcal{A}_{\gamma}(d) = \frac{d - 3}{\sqrt{(d - 2)(d - 4)}} \sin\left(\gamma \arctan\left(\sqrt{(d - 2)(d - 4)}\right) \right).
\end{equation}
This trigonometric formula shows that $\mathcal{A}_{\gamma}(d)$ is always real despite $r$ being complex for all $d > 4$. It is also an odd function of $\gamma$, as is expected from \eqref{A formula}.

Similarly, the generating function for $\mathcal{B}_{n}$ is
\begin{equation}\label{B Gen f'n}
\sum_{n=0}^{\infty} \mathcal{B}_{n}(q) t^{n} = \frac{q d}{(1 - qt + t^{2})(1 - t)}.
\end{equation}
Expanding \eqref{B Gen f'n} as before in a power series, we find that
\begin{equation}\label{B formula}
\mathcal{B}_{n}(q) = \frac{q \left(r - r^{-1} + r^{n} - r^{-n} - r^{n+1} + r^{-n-1}\right)}{(r - r^{-1})(r - 1)(r^{-1} - 1)},
\end{equation}
where $r$ is defined as before. From this, it is easy to see that replacing $n \rightarrow -n-1$ leaves $\mathcal{B}_{n}(q)$ invariant, as claimed above. We take the same continuous generalization as before, replacing $n \in \mathbb{Z}$ with $\gamma \in \mathbb{R}$, and note that this must indeed be the same continuous parameter that indexes the $\mathcal{A}_{\gamma}$ for the identifications with $n$ to hold. Again, a trigonometric formula can be obtained using a computer algebra system:
\begin{multline}\label{B trigonometric}
\mathcal{B}_{\gamma}(d) = \frac{1}{d-4} \Bigg(1 - \cos\left(\gamma \arctan\left(\sqrt{(d - 2)(d - 4)}\right) \right) \\
+ \frac{d - 4}{\sqrt{(d - 2)(d - 4)}} \sin\left(\gamma \arctan\left(\sqrt{(d - 2)(d - 4)}\right) \right)\Bigg).
\end{multline}
Thus, with \eqref{A trigonometric} and \eqref{B trigonometric}, the resulting $d$-dimensional metric
\begin{equation}\label{Continuous Transformation}
ds^{2} = \pm g_{11}^{\mathcal{A}_{\gamma+1}(d)} g_{22}^{\mathcal{A}_{\gamma}(d)} \, dx^{2} \pm g_{11}^{-\mathcal{A}_{\gamma}(d)} g_{22}^{-\mathcal{A}_{\gamma-1}(d)} \, dy^{2} + g_{11}^{\mathcal{B}_{\gamma}(d)} g_{22}^{\mathcal{B}_{\gamma-1}(d)} g_{ij} \, dz^{i} dz^{j},
\end{equation}
is Ricci-flat by construction for any $\gamma\in \mathbb{R}$. To the best of our knowledge, this is the first known example of a continuous Lie symmetry used for solution generation in higher-dimensional vacuum Einstein equations.\footnote{One can verify that the explicit formulas for $\mathcal{A}_{\gamma}(d)$ and $\mathcal{B}_{\gamma}(d)$ satisfy the recursion relation and the initial conditions using sum-to-product trigonometric identities. These trigonometric formulas demonstrate that $\mathcal{A}_{\gamma}(d)$ and $\mathcal{B}_{\gamma}(d)$ are real for $d\ge 5$, just as they are for  $d=4$. }

For $d=4$, the family \eqref{Continuous Transformation} becomes (equivalent to equation (3.14) in \cite{AkbarMacCallum})
\begin{equation}\label{4d General}
ds^{2} = \pm g_{11}^{\gamma+1}g_{22}^{\gamma} dx^{2} + g_{11}^{-\gamma}g_{22}^{-\gamma + 1}dy^{2} + g_{11}^{\gamma(\gamma+1)}g_{22}^{\gamma(\gamma-1)}g_{ij}dz^{i}dz^{j}.
\end{equation}
For $d=5$, one obtains
\begin{multline}\label{5d General}
ds^{2} = \pm g_{11}^{\frac{2}{\sqrt{3}}\sin(\frac{\pi(2\gamma+1)}{3})}g_{22}^{\frac{2}{\sqrt{3}}\sin(\frac{2\pi \gamma}{3})}dx^{2} + g_{11}^{-\frac{2}{\sqrt{3}}\sin(\frac{2\pi \gamma}{3})}g_{22}^{-\frac{2}{\sqrt{3}}\sin(\frac{\pi(2\gamma-1)}{3})}dy^{2}\\
 + g_{11}^{1-\cos(\frac{2\pi \gamma}{3}) + \frac{1}{\sqrt{3}}\sin(\frac{2\pi \gamma}{3})}g_{22}^{1-\cos(\frac{\pi(2\gamma-1)}{3} + \sin(\frac{\pi(2\gamma-1)}{3})}g_{ij}dz^{i}dz^{j}.
\end{multline}
General formulas for the $\mathcal{A}_{\gamma}$ and $\mathcal{B}_{\gamma}$ in different dimensions are tabulated in Table 1. It is useful to see how this works in a somewhat more familiar context. The four and five-dimensional Schwarzschild metrics, both of which can be written in the warped form with $t$ and $\phi$ as the two static coordinates, respectively, give
\begin{align}\label{4d Schwarszchild Transformed}
 ds^{2} = & - \left(1-\frac{2m}{r}\right)^{\gamma+1}(r\sin\theta)^{2\gamma}dt^{2} + \left(1-\frac{2m}{r}\right)^{-\gamma}(r\sin\theta)^{-2\gamma+2}d\phi^{2} \nonumber \\
 & + \left(1-\frac{2m}{r}\right)^{\gamma(\gamma+1)}(r\sin\theta)^{2\gamma(\gamma-1)}\left[\left(1-\frac{2m}{r}\right)^{-1}dr^{2} + r^{2}d\theta^{2}\right]
\end{align}
and
\begin{align}\label{5d Schwarszchild Transformed}
ds^{2} = & -\left(1-\frac{2m}{r^{2}}\right)^{\frac{2}{\sqrt{3}}\sin(\frac{\pi(\gamma+1)}{3})}(r\sin\theta \sin\chi)^{\frac{4}{\sqrt{3}}\sin(\frac{\pi \gamma}{3})}dt^{2} \nonumber \\ & + \left(1-\frac{2m}{r^{2}}\right)^{-\frac{2}{\sqrt{3}}\sin(\frac{\pi \gamma}{3})}(r \sin\theta \sin\chi)^{-\frac{4}{\sqrt{3}}\sin(\frac{\pi(\gamma-1)}{3})}d\phi^{2} \\ & + \left(1-\frac{2m}{r^{2}}\right)^{1-\cos(\frac{\pi \gamma}{3}) + \frac{1}{\sqrt{3}}\sin(\frac{\pi \gamma}{3})}(r \sin\theta \sin\chi)^{2[1-\cos(\frac{\pi(\gamma-1)}{3}) + \frac{1}{\sqrt{3}}\sin(\frac{\pi(\gamma-1)}{3})]} \nonumber \\
& \left[\left((1-\frac{2m}{r^{2}}\right)^{-1}dr^{2} + r^{2}(d\theta^{2} + \sin^{2}\theta d\chi^{2})\right]. \nonumber
\end{align}
Metric \eqref{4d Schwarszchild Transformed} appeared in \cite{AkbarMacCallum}, while \eqref{5d Schwarszchild Transformed} is new. According to the above result, both are Ricci flat for all $\gamma \in \mathbb{R}$, and as one can check with a computer algebra system.

\begin{table}[h!]
\centering
\begin{tabular}{|c|c|c|}
\hline
$d$ & $\mathcal{A}_{\gamma}$ & $\mathcal{B}_{\gamma}$ \\
\hline
$4$ & $\gamma$ & $\gamma(\gamma+1)$ \\
$5$ & $\frac{2}{\sqrt{3}}\sin\left(\frac{\pi \gamma}{3}\right)$ & $1 - \cos\left(\frac{\pi \gamma}{3}\right) + \frac{1}{\sqrt{3}} \sin\left(\frac{\pi \gamma}{3}\right)$ \\
$6$ & $\frac{3}{2\sqrt{2}} \sin\left(\gamma \arctan(2\sqrt{2})\right)$ & $\frac{1}{2} \left(1 - \cos\left(\gamma \arctan(2\sqrt{2})\right) + \frac{1}{\sqrt{2}} \sin\left(\gamma \arctan(2\sqrt{2})\right)\right)$ \\
$7$ & $\frac{4}{\sqrt{15}} \sin\left(\gamma \arctan(\sqrt{15})\right)$ & $\frac{1}{3} \left(1 - \cos\left(\gamma \arctan(\sqrt{15})\right) + \sqrt{\frac{3}{5}} \sin\left(\gamma \arctan(\sqrt{15})\right)\right)$ \\
$8$ & $\frac{5}{2\sqrt{6}} \sin\left(\gamma \arctan(2\sqrt{6})\right)$ & $\frac{1}{4} \left(1 - \cos\left(\gamma \arctan(2\sqrt{6})\right) + \sqrt{\frac{2}{3}} \sin\left(\gamma \arctan(2\sqrt{6})\right)\right)$ \\
\hline
\end{tabular}
\caption{$\mathcal{A}{\gamma}$ and $\mathcal{B}{\gamma}$ in $d\ge 4$ dimensions. They have a polynomial form in four dimensions due to the presence of the double root. In dimension five, they can be written in slightly simpler trigonometric form using identities.}
\end{table}
\subsection{Group Properties}\label{sec:group}
Let us denote the transformation that maps the seed metric \eqref{Base Metric} into \eqref{Continuous Transformation} as $T_{\gamma}$. The combined action of $T_{\beta} \circ T_{\alpha}$ takes metric \eqref{Base Metric} into
\begin{eqnarray}\label{Tbeta o Talpha}
ds^{2} &=& \pm g_{11}^{\mathcal{A}_{2\alpha+1}(d) \mathcal{A}_{2\beta+1}(d) - \mathcal{A}_{2\alpha}(d) \mathcal{A}_{2\beta}(d)} g_{22}^{\mathcal{A}_{2\alpha}(d) \mathcal{A}_{2\beta+1}(d) - \mathcal{A}_{2\alpha-1}(d) \mathcal{A}_{2\beta}(d)} \, dx^{2} \nonumber \\
&& \pm g_{11}^{\mathcal{A}_{2\alpha}(d) \mathcal{A}_{2\beta-1}(d) - \mathcal{A}_{2\alpha+1}(d) \mathcal{A}_{2\beta}(d)} g_{22}^{\mathcal{A}_{2\alpha-1}(d) \mathcal{A}_{2\beta-1}(d) - \mathcal{A}_{2\alpha}(d) \mathcal{A}_{2\beta}(d)} \, dy^{2} \nonumber \\
&& + g_{11}^{\mathcal{B}_{2\beta}(d) \mathcal{A}_{2\alpha+1}(d) - \mathcal{B}_{2\beta-1}(d) \mathcal{A}_{2\alpha}(d)} g_{22}^{\mathcal{B}_{2\beta}(d) \mathcal{A}_{2\alpha}(d) - \mathcal{B}_{2\beta-1}(d) \mathcal{A}_{2\alpha-1}(d)} g_{ij} \, dz^{i} dz^{j}.
\end{eqnarray}
We can verify the following identities
\begin{align}
\mathcal{A}_{2\alpha+1}(d) \mathcal{A}_{2\beta+1}(d) - \mathcal{A}_{2\alpha}(d) \mathcal{A}_{2\beta}(d) &= \mathcal{A}_{2\alpha+2\beta+1}(d), \nonumber \\
\mathcal{B}_{2\beta}(d) \mathcal{A}_{2\alpha+1}(d) - \mathcal{B}_{2\beta-1}(d) \mathcal{A}_{2\alpha}(d) &= \mathcal{B}_{2\alpha+2\beta}(d).
\end{align}
With these identities, as well as the symmetry $\mathcal{A}_{-\gamma}(d) = -\mathcal{A}_{\gamma}(d)$ shown earlier, it is clear that \eqref{Tbeta o Talpha} is simply the transformation $T_{\beta + \alpha}$ applied to the seed metric \eqref{Base Metric}. Hence, the set of transformations is closed under composition, inheriting the identity element, $T_{0} = \mathbb{I}$, and inverses, $T_{\alpha} T_{-\alpha} = T_{0} = \mathbb{I}$, from the isomorphism with addition on the real line. Thus, the set of transformations forms a group, representing a Lie point symmetry on the set of Ricci-flat metrics in the arbitrary-dimensional case, just as in four dimensions \cite{AkbarMacCallum}.

\section{Ricci Solitons}
Ricci flow is an intrinsic geometric flow in which the metric $g_{\mu\nu}$ on a manifold $M^{d}$ evolves by its Ricci curvature tensor:
\begin{equation}
\frac{\partial g_{\mu\nu}}{\partial \eta} = -2R_{\mu\nu}.
\label{eq1.1}
\end{equation}
This plays an important role in geometric analysis and also arises in physics (readers are referred to standard references for more details \cite{CK, Chow1}). Ricci solitons are self-similar solutions of this flow in which the metric only evolves by scaling and diffeomorphism. The initial metric of a Ricci soliton can be shown to satisfy the following equation \cite{CK, Chow1}
\begin{equation}
R_{\mu\nu}-\frac12{\cal L}_X g_{\mu\nu}=\kappa g_{\mu\nu} \label{eq1.5}
\end{equation}
where $X$ is the vector field generating the diffeomorphism.

Combining the well-known correspondence between static metrics and Einstein-scalar field theory in one lower dimension with the correspondence between Ricci solitons and the Einstein-scalar system, also in one lower dimension, as found in \cite{AkbarWoolgar}, one can obtain a Ricci soliton in $d$ dimensions from any static $d$-dimensional Ricci-flat metric. For an explicit continuous symmetry of the latter, one obtains a one-parameter family of Ricci solitons, just as one obtains a one-parameter family of Ricci-flat metrics, without having to solve any equations. For the Lie point symmetry at hand, the four-dimensional Ricci soliton metrics were worked out in \cite{AkbarMacCallum} in Weyl coordinates. Here, we obtain the soliton metric in the warped product form (\ref{Base Metric}) for $d\ge 4$.

More precisely, every $d$-dimensional static Ricci-flat metric of the form
\begin{equation}
ds^2 =\pm e^{2u}dt^2+e^{-\frac{2u}{d-3}}\tilde{g}_{ij}dx^idx^j\label{dRicciflat}
\end{equation}
is in one-to-one correspondence with the following Ricci soliton metric in $d$ dimensions
\begin{equation}
ds^2 =e^{2\sqrt{\frac{d-2}{d-3}}u}dt^2+\tilde{g}_{ij}dx^idx^j\label{ndriccisol}
\end{equation}
which solves the Ricci soliton equation (\ref{eq1.5}) with $\kappa=0$ (i.e., it is a ``steady" soliton), where
\begin{equation} \label{diffeofield}
X:=- 2\sqrt{\frac{d-2}{d-3}}\, \tilde{g}^{ij}\tilde{\nabla}_i u \frac{\partial}{\partial x^j}.
\end{equation}
Writing the metric (\ref{Base Metric}) in this form and working through the algebra, we find that for every warped product Ricci-flat metric of the form (\ref{Base Metric}) one obtains the following Ricci soltion
\begin{eqnarray}
ds^{2}& =& \pm g_{11}^{\sqrt{\frac{d-2}{d-3}}\mathcal{A}_{\gamma+1}(d)} g_{22}^{\sqrt{\frac{d-2}{d-3}}\mathcal{A}_{\gamma}(d)} \, dx^{2} \pm g_{11}^{-\frac{\mathcal{A}_{\gamma}(d)}{d-3}} g_{22}^{-\frac{\mathcal{A}_{\gamma-1}(d)}{d-3}} \, dy^{2} \nonumber \\
&& + g_{11}^{\mathcal{B}_{\gamma}(d)-\frac{\mathcal{A}_{\gamma+1}(d)}{d-3}} g_{22}^{\mathcal{B}_{\gamma-1}(d)-\frac{\mathcal{A}_{\gamma}(d)}{d-3}} g_{ij} \, dz^{i} dz^{j}\nonumber\\
&\equiv&  \pm g_{11}^{\sqrt{\frac{d-2}{d-3}}\mathcal{A}_{\gamma+1}(d)} g_{22}^{\sqrt{\frac{d-2}{d-3}}\mathcal{A}_{\gamma}(d)} \, dx^{2}+\tilde{g}_{ij}dx^idx^j.
\end{eqnarray}
where the vector field $X$ in equation (\ref{diffeofield}) is to be computed from the co-dimension one part of the metric $\tilde{g}_{ij}$ above with
\begin{equation}
u=\frac{1}{2}\left[\mathcal{A}_{\gamma+1}(d) \ln(g_{11})+\mathcal{A}_{\gamma}(d) \ln(g_{22})\right].
\end{equation}
From these, explicit examples can easily be worked by the interested reader. The only computationally lengthy part is determining the vector field $X$ explicitly, but this is straightforward.

\section{Conclusion}
In four dimensions, the presence of Lie point symmetries in the vacuum Einstein equations is quite apparent when viewed through Weyl coordinates. They stem from the linearity of Laplace’s equation --- superposing any ``Newtonian" potential onto the potential of any solution leads to a one-parameter generalization of the latter. However, only in the special case when the superposing potential is \(\ln\rho\) one obtains an explicit algebraic prescription that can act on any metric and generate a one-parameter family of solutions distinct from the original seed \cite{AkbarMacCallum}.

In this paper, we found that the above Lie point symmetry extends to arbitrary dimensions. Although Weyl coordinates do not exist in the presence of only two Killing vector fields in $d\ge5$, using the doubly-warped product form of the metric (\ref{Base Metric}) for arbitrary dimensions, we were able to build this continuous symmetry off the discrete Buchdahl transformation. This symmetry exists for any value of the (real) parameter and apprears as nontrivial exponents of the metric components. However, unlike in four dimensions, they are non-polynomial for all $d\ge 5$ but can still be expressed in a universal trigonometric form. Even with the most advanced computational methods available today, it is unlikely that one could directly derive the one-parameter solutions from Einstein’s equations in any feasible way. At the same time, it would be rather challenging to spot this symmetry by studying the vacuum Einstein equations using standard symmetry methods in differential equations (see \cite{Olver, Stephani}).

In the presence of more than two hypersurface-orthogonal Killing vector fields in the vacuum, multiple sets of Killing vector pairs are possible, and each pair will give rise to a distinct one-parameter Lie group and a distinct one-parameter generalization of the seed metric. For instance, in the case of three Killing vector fields, there would be three one-parameter Lie groups, which would generally be non-commuting, each with its own distinct generalization. It would be interesting to explore whether these group actions can be meaningfully combined, a question left for future work.

\section*{Acknowledgement}
We thank Behshid Kasmaie for useful discussions. MS acknowledges support from NSF REU Program NSF-2057810 (PI Lindsay King) and the hospitality of the University of Texas at Dallas.

\end{document}